\documentclass[a4paper]{aa}

\usepackage{graphicx,natbib}
\usepackage{txfonts}
\usepackage[usenames,dvipsnames]{xcolor}

\newcommand{\bz}{$\langle B_{\rm z} \rangle$}
\newcommand{\bb}{$\langle B \rangle$}
\newcommand{\bq}{$\langle B_{\rm q} \rangle$}
\newcommand{\bt}{$\langle B_{\rm t} \rangle$}
\newcommand{\vs}{$v_{\rm e}\sin i$}
\newcommand{\kms}{km\,s$^{-1}$}

\newcommand{\teff}{$T_{\rm eff}$}
\newcommand{\lgg}{$\log g$}

\newcommand{\figps}[1]{\resizebox{\hsize}{!}{\rotatebox{0}{\includegraphics{#1}}}}

\newcommand{\fifps}[2]{\centering\resizebox{#1}{!}{\includegraphics{#2}}}

\begin{document}

\title{Are there tangled magnetic fields on HgMn stars?%
\thanks{Based on observations collected at the European Southern Observatory, Chile (ESO programmes 084.D-0338, 085.D-0296, 086.D-0240).}}

\author{O.~Kochukhov\inst{1}
   \and V.~Makaganiuk\inst{1}
   \and N.~Piskunov\inst{1}
   \and S.~V.~Jeffers\inst{2}
   \and C.~M.~Johns-Krull\inst{3}
   \and C.~U.~Keller\inst{4} 
   \and \\ M.~Rodenhuis\inst{4}
   \and F.~Snik\inst{4}
   \and H.~C.~Stempels\inst{1}
   \and J.~A.~Valenti\inst{5}}

\institute{Department Physics and Astronomy, Uppsala University, Box 516, 751 20 Uppsala, Sweden
\and
Institute of Astrophysics, Georg-August University, Friedrich-Hund-Platz 1, D-37077 G\"ottingen, Germany
\and
Department of Physics and Astronomy, Rice University, 6100 Main Street, Houston, TX 77005, USA
\and
Sterrewacht Leiden, Universiteit Leiden, Niels Bohrweg 2, 2333 CA Leiden, The Netherlands
\and
Space Telescope Science Institute, 3700 San Martin Dr, Baltimore MD 21211, USA}

\date{Received 13 March 2013 / Accepted 22 April 2013}

\authorrunning{O. Kochukhov et al.}

\abstract
{Several recent spectrophotometric studies failed to detect significant global magnetic fields in late-B HgMn chemically peculiar stars, but some investigations have suggested the presence of strong unstructured or tangled fields in these objects.}
{We used detailed spectrum synthesis analysis to search for evidence of tangled magnetic fields in high-quality observed spectra of 8 slowly rotating HgMn stars and one normal late-B star. We also evaluated recent sporadic detections of weak longitudinal magnetic fields in HgMn stars based on the moment technique.}
{Our spectrum synthesis code calculated the Zeeman broadening of metal lines in HARPS spectra, assuming an unstructured, turbulent magnetic field. A simple line formation model with a homogeneous radial field distribution was applied to assess compatibility between previous longitudinal field measurements and the observed mean circular polarization signatures.}
{Our analysis of the Zeeman broadening of magnetically sensitive spectral lines reveals no evidence of tangled magnetic fields in any of the studied HgMn or normal stars. We infer upper limits of 200--700~G for the mean magnetic field modulus -- much smaller than the field strengths implied by studies based on differential magnetic line intensification and quadratic field diagnostics. The new HARPSpol longitudinal field measurements for the extreme HgMn star HD\,65949 and the normal late-B star 21\,Peg  are consistent with zero at a precision of 3--6~G. Re-analysis of our Stokes $V$ spectra of the spotted HgMn star HD\,11753 shows that the recent moment technique measurements retrieved from the same data are incompatible with the lack of circular polarization signatures in the spectrum of this star.}
{We conclude that there is no evidence for substantial tangled magnetic fields on the surfaces of studied HgMn stars. We cannot independently confirm the presence of very strong quadratic or marginal longitudinal fields for these stars, so results from the moment technique are likely to be spurious.}

\keywords{stars: atmospheres -- stars: chemically peculiar -- stars: individual: HD\,35548, HD\,65949, HD\,71066, HD\,78316, HD\,175640, HD\,178065, HD\,186122, HD\,193452, HD\,209459 -- stars: magnetic field -- polarization}

\maketitle

\section{Introduction}
\label{intro}

Mercury-manganese (HgMn) stars comprise a group of late-B chemically peculiar stars distinguished by a strong overabundance and an unusual isotopic composition of heavy elements \citep{adelman:2004,woolf:1999}. Many of these stars are slow rotators and members of close spectroscopic binary systems \citep{abt:2002,catanzaro:2004}. The properties of HgMn stars facilitate precise chemical abundance analysis of their atmospheres \citep[e.g.][]{adelman:2006}, making them the preferred targets for detailed comparisons of observed surface abundance patterns and predictions from atomic diffusion theory \citep{michaud:1974,alecian:1981a}.

The unexpected discovery of low-contrast abundance inhomogeneities on the surfaces of HgMn stars has rekindled interest in the astrophysical processes that operate in these stars \citep{adelman:2002,kochukhov:2005b,hubrig:2006,kochukhov:2011b}. Chemical spots are often found in magnetic B-type (Bp) stars, which overlap with HgMn stars on the H-R diagram. Bp stars possess strong global magnetic fields, which are believed to be responsible for the chemical spot formation \citep[e.g.][]{michaud:1981}. In contrast to this clear link between magnetic fields and chemical spots in Bp stars, no robust and reproducible magnetic field detections have ever been reported for any of the HgMn stars. Moreover, inhomogeneities on HgMn stars evolve with time \citep{kochukhov:2007b,briquet:2010}, possibly indicating a previously unknown time-dependent, non-equilibrium diffusion process \citep{alecian:2011}. This behavior is, again, very different from that of spots on magnetic Bp stars, which are observed to be stable for at least several decades \citep{adelman:2001a}.

The possible role of weak magnetic fields in explaining the puzzling surface phenomena observed in HgMn stars is a topic of ongoing debate. The literature contains claims of magnetic field detections based on low- and high-resolution spectropolarimetric observations \citep{hubrig:2006b,hubrig:2010,hubrig:2012}. However, many other detailed circular polarization studies of individual spotted HgMn stars revealed no magnetic field signatures in spectral line profiles \citep{wade:2006,folsom:2010,makaganiuk:2011,makaganiuk:2012,kochukhov:2011b}. Similar null results were also reported by several spectropolarimetric surveys that included a large number of HgMn-type stars \citep{shorlin:2002,auriere:2010a,makaganiuk:2011a}. These studies established an upper limit from a few tens of G to just a few G for the mean longitudinal magnetic field in HgMn stars.

The failure of spectropolarimetric studies to unambiguously detect the fields in HgMn stars may be attributed to the complexity of the surface magnetic field topologies \citep[e.g.][]{hubrig:1998}. Indeed, spectropolarimetry and, especially, the mean longitudinal magnetic field diagnostic is only sensitive to the line-of-sight magnetic field component. Theoretically, the net line-of-sight magnetic field can be close to zero if many regions of opposite polarity are present on the stellar surface. Highly structured magnetic field topologies can reduce the Stokes $V$ signatures in spectral line profiles below the detection threshold. On the other hand, field orientation has a relatively minor effect on magnetic broadening and splitting, so this diagnostic can reveal complex magnetic fields invisible to spectropolarimetry. This type of analysis has yielded a number of surprisingly high magnetic field estimates for HgMn stars. For instance, on the basis of studying the differential magnetic intensification of \ion{Fe}{ii} lines, \citet{hubrig:1999a} and \citet{hubrig:2001} concluded that some HgMn stars possess complex or ``tangled'' magnetic fields stronger than 2~kG. A complementary quadratic magnetic field diagnostic method based on the comparison of magnetic broadening in spectral lines with different Zeeman splitting patterns suggested the presence of even stronger, 2--8~kG, magnetic fields in HgMn stars \citep{mathys:1995b,hubrig:2012}.

Both the magnetic intensification and quadratic field diagnostic methods rely on a number of simplifying assumptions about the spectral line formation in magnetic field. These methods were originally developed for intensity spectra with limited wavelength coverage and for spectropolarimetric observations with moderate resolution and relatively low signal-to-noise (S/N) ratio. Modern \'echelle spectra of HgMn stars provide access to numerous magnetically sensitive lines that can be analyzed using sophisticated theoretical tools. The goal of our paper is to probe the existence of complex magnetic fields in HgMn stars by taking advantage of the high-resolution, high S/N observations and using accurate methods to model magnetic radiative transfer. We also critically examine spectropolarimetric magnetic field detections that seem to contradict our null results for a few HgMn stars \citep{makaganiuk:2011,makaganiuk:2011a,makaganiuk:2012}.

The rest of the paper is organized as follows. Sect.~\ref{methods} presents our spectroscopic and spectropolarimetric observations and discusses corresponding analysis methods. Results of the magnetic field search using circular polarization measurements, differential magnetic intensification, and Zeeman broadening of spectral lines are reported in Sect.~\ref{results}. Finally, Sect.~\ref{discussion} compares our results with the outcome of previous investigations and assesses implications of our study for the general question of the presence of magnetic fields in HgMn stars.

\section{Methods}
\label{methods}

\subsection{Observations and data reduction}
\label{obs}

We analyze spectra of HgMn stars obtained with the circular polarimetric mode \citep{snik:2011,piskunov:2011} of the HARPS spectrometer \citep{mayor:2003} installed at the 3.6-m ESO telescope in La Silla. Most observations were obtained during several observing runs in 2010. The acquisition and reduction of these data was described in detail by \citet{makaganiuk:2011a,makaganiuk:2011,makaganiuk:2012}. We also obtained new circular polarization observations of the HgMn star HD\,65949 and analyzed spectra of the normal late-B star HD\,209459, which was observed together with the HgMn stars from our sample but was not studied by \citeauthor{makaganiuk:2011a}

The HARPSpol instrument records spectra in the wavelength range 3780--6913~\AA\ with an 80 \AA\ gap around 5300~\AA. All stars in our study were observed using the circular polarization analyzer. Observations of each star were typically split into four sub-exposures between which the quarter-wave retarder plate was rotated in 90\degr\ steps relative to the beamsplitter. The spectra were extracted and calibrated using a dedicated version of the {\sc reduce} pipeline \citep{piskunov:2002}. The Stokes $I$ spectrum was obtained by averaging the right- and left-hand polarized spectra from all sub-exposures of a given star. The Stokes $V$ and diagnostic null spectra were derived by combining the extracted spectra according to the ``ratio'' spectropolarimetric demodulation method \citep{donati:1997,bagnulo:2009}. Analysis of the emission lines in the ThAr comparison spectra showed that the resolving power of HARPS in polarimetric mode is $R=\lambda/\Delta\lambda=109\,000$ with about 1--2\% variation across the \'echelle format.

Subtle Zeeman broadening due to a weak and complex magnetic field can be reliably detected only for very slowly rotating stars. Therefore, we selected a small number of sharp-lined targets out of the full sample of HgMn stars examined in the survey by \citet{makaganiuk:2011a}. Our primary targets include 7 HgMn stars with \vs\,$\le$\,2~\kms: HD\,35548, HD\,65949, HD\,71066, HD\,175640, HD\,178065, HD\,186122, and HD\,193452. For comparison, we also analyzed two stars with projected rotational velocities in the range of 4--6~\kms: HD\,78316 ($\kappa$~Cnc) and HD\,209459 (21\,Peg). Table~\ref{tab1} summarizes alternative designations, atmospheric parameters and literature information on the rotational velocities of our program stars.

Preliminary continuum normalization of Stokes $I$ spectra was performed by consecutively applying the blaze function, an empirical response function, and smooth fit corrections as described by \citet{makaganiuk:2011a}. The resulting continuum normalization has a precision of approximately 0.5\%. The final zeroth-order correction to the continuum was applied as part of the modeling of intensity profiles of individual spectral lines described below (Sect.~\ref{bturb}).

Characteristics of the HARPSpol observations of 9 program stars are provided in Table~\ref{tab2}. We list the UT dates of individual observations, the corresponding heliocentric Julian dates, the total exposure time, and the peak S/N per pixel measured at $\lambda\approx5200$~\AA. Two observations were obtained for 21\,Peg. Both were used to measure the mean longitudinal magnetic field (Sect.~\ref{nobz}), but only the higher quality one was employed for the Stokes $I$ Zeeman broadening analysis (Sect.~\ref{bturb}).

\subsection{Multi-line polarization analysis}
\label{lsd}

A search for weak magnetic field signatures in high-resolution stellar Stokes $V$ spectra is greatly facilitated by a multi-line analysis. In this study we used the least-squares deconvolution (LSD) code developed by \citet{kochukhov:2010a}. The LSD method, originally introduced by \citet{donati:1997}, combines information from all suitable stellar absorption lines under the assumption of self-similarity of the line profile shapes and a linear addition of overlapping lines. \citet{kochukhov:2010a} showed that these assumptions are adequate for weak and moderately strong magnetic fields and that the mean longitudinal magnetic field inferred from the Stokes $V$ LSD profiles is not affected by systematic biases.

In this study we applied the LSD analysis to our new spectropolarimetric observations of HD\,65949 and HD\,209459. For the remaining targets the LSD longitudinal field measurements were published by \citet{makaganiuk:2011a}. For completeness, Table~\ref{tab2} reproduces these measurements. The line lists necessary for applying LSD were obtained from the VALD database \citep{kupka:1999}, using atmospheric parameters listed in Table~\ref{tab1} and adopting chemical abundances from the studies by \citet{cowley:2010} and \citet{fossati:2009} for HD\,65949 and HD\,209459, respectively. In total, we used 500--600 lines deeper than 10\% of the continuum for each star. The application of LSD increased the S/N of the Stokes $V$ profiles by a factor of $\approx$\,6. Sect.~\ref{nobz} describes the analysis of the LSD profiles of HD\,65949 and HD\,209459.

\subsection{Magnetic intensification and broadening of spectral lines}
\label{imodel}

The Zeeman effect produces intensification, broadening, and splitting of a spectral line depending on its effective Land\'e factor and the magnetic splitting pattern. However, unless the field is very strong and the Zeeman splitting can be resolved and measured directly, there are few, if any, reliable model-independent methods to derive magnetic field intensity from the Stokes $I$ spectra. In particular, as we will show below, such techniques as differential line intensification and quadratic field diagnostic are prone to biases and lead to contradictory results for late-B stars. In this situation, detailed spectrum synthesis is the only robust way to detect and measure magnetic fields with strengths below 1--2~kG \citep{kochukhov:2002,kochukhov:2006b,johns-krull:2007}. Refined analysis of high-resolution spectra of bright stars can even reveal fields in the 400--500~G range \citep{anderson:2010}.

Here we use the {\sc synmast} magnetic spectrum synthesis code \citep{kochukhov:2010a} to simulate the effects of different magnetic field geometries on the profiles of magnetically sensitive metal lines. This code calculates the four Stokes parameter spectra for a homogeneous magnetic field distribution characterized by the three vector components specified in the stellar coordinate system \citep[see][]{kochukhov:2007d} or for a turbulent magnetic field as described below. These calculations are based on the {\sc llmodels} model atmospheres \citep{shulyak:2004}, computed for the stellar parameters given in Table~\ref{tab1} taking into account individual abundances of the program stars. We set the microturbulent velocity to zero, since HgMn stars are known to lack signatures of turbulence in their atmospheres \citep{landstreet:2009}.

\begin{figure}[!th]
\centering
\figps{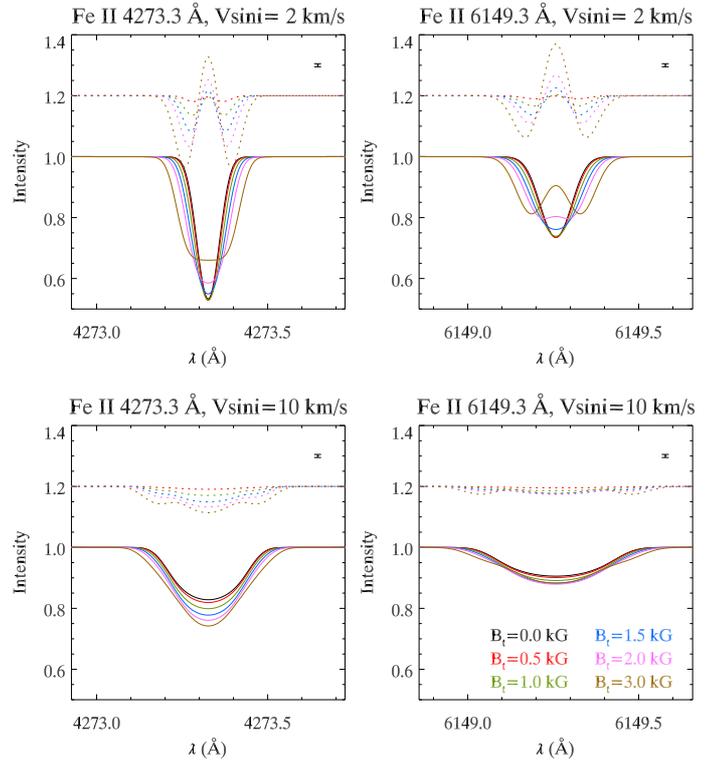}
\caption{Effect of the turbulent magnetic field on the \ion{Fe}{ii} 4273.3 and 6149.3~\AA\ lines. The solid curves show synthetic profiles for the field strengths ranging from 0 to 3 kG and for \vs\,=\,2 and 10~\kms. The dotted lines show the difference between the magnetic and non-magnetic profiles, offset vertically by 1.2. The error bar in the upper right corner of each panel corresponds to the S/N ratio of 200:1.}
\label{fig:syn_lines}
\end{figure}

As mentioned in the introduction, previous studies of HgMn stars could not unambiguously detect Zeeman-induced polarization in spectral lines. Even if we believe a few disputed detections, the polarimetrically-inferred magnetic fields in HgMn stars typically do not exceed $\sim$\,100~G. At the same time, multi-kG fields were inferred from the analysis of intensity spectra of the same or similar stars. These vastly different field strength estimates can be reconciled only if the fields are arranged on the stellar surfaces in a highly structured, complex configuration. To model such fields we use the concept of an isotropic, turbulent magnetic field \citep{polarization:2004}, which implies magnetic field parameters that are uncorrelated on the characteristic scale of radiative transfer. A turbulent field represents an opposite extreme compared to an organized, large-scale magnetic field topology. Yet, it is the only known magnetic field model satisfying the observational requirement of giving a noticeable Zeeman broadening and intensification in the intensity spectra without producing a detectable circular polarization.

In general, the isotropic, turbulent magnetic field is characterized by some distribution of the magnetic field modulus, $F(B_{\rm t})$. For this magnetic field topology, the polarized radiative transfer equation terms that depend on the field orientation average to zero, yielding null emergent $QUV$ spectra but retaining the magnetic field effects in Stokes $I$. We implemented this model in the {\sc synmast} code assuming, as a first approximation, that the entire stellar surface is covered by the magnetic field with a strength \bt. The disk-integrated spectra for arbitrary \vs\ are then produced from the intensity calculations at several limb angles, similar to a non-magnetic spectrum synthesis \citep{kochukhov:2007d}.

Figure~\ref{fig:syn_lines} illustrates the effect of a turbulent magnetic field on the \ion{Fe}{ii} 4273.3 ($z=2.15$, pseudo-triplet splitting) and 6149.3~\AA\ ($z=1.35$, doublet splitting) spectral lines. These calculations were done for the \teff\,=\,12000~K, \lgg\,=\,4.0 model atmosphere and a spectral resolution $R=109\,000$. As one can see from this figure, for a very low \vs, both lines show detectable profile distortions for \bt\,$\ge$\,0.5--1.0~kG fields. Starting from $\sim$\,2~kG, the partially-resolved Zeeman splitting becomes obvious in the \ion{Fe}{ii} 6149.3~\AA\ line. For higher projected rotational velocities a stronger field is required to produce significant line profile distortions. Based on these calculations, we conclude that, provided \vs\ can be independently estimated from magnetically insensitive lines, it is possible to detect kG-strength tangled magnetic fields using magnetically sensitive spectral lines in S/N\,$\ge$\,200, $R$\,$\ge$\,$10^5$ observations of slowly rotating late-B stars.

\subsection{Modeling of circular polarization profiles}
\label{vmodel}

Historically, the efforts to detect magnetic fields in early-type chemically peculiar stars mostly relied on the longitudinal field diagnostic \citep[e.g.][]{borra:1980,mathys:1991}. However, modern availability of the high-quality mean Stokes $V$ profiles enables several other field detection and assessment methods. For instance, \citet{petit:2012} demonstrated how a useful estimate of the surface dipolar field strength can be obtained from the LSD Stokes $V$ profiles without complete rotational phase coverage. \citet{shultz:2012} used a similar polarization modeling approach to test compatibility of their high-resolution Stokes $V$ observations with the magnetic field models derived solely from the \bz\ measurements. Here we apply a similar technique to assess the circular polarization signatures expected for individual \bz\ detections reported in the literature.

To model the observed LSD Stokes $I$ and $V$ profiles, we use the polarization spectrum synthesis method described by \citet{petit:2012}. The local intensity profile is represented by a Gaussian function while the Stokes $V$ is given by the scaled derivative of the Gaussian according to the well-known weak-field approximation formula \citep{polarization:2004}. The wavelength and effective Land\'e factor required by this approximation are given by the respective mean values for the LSD mask. The Stokes $I$ and $V$ spectra are obtained by numerical integration of these local profiles over the visible stellar hemisphere for a given magnetic field geometry. For the calculations presented in this paper, we adopted a homogeneous radial field distribution with the strength adjusted to match a particular \bz\ value. Other free parameters of the model (\vs\ and line depth) were obtained by fitting the observed Stokes $I$ profile.

\begin{figure}[!th]
\centering
\figps{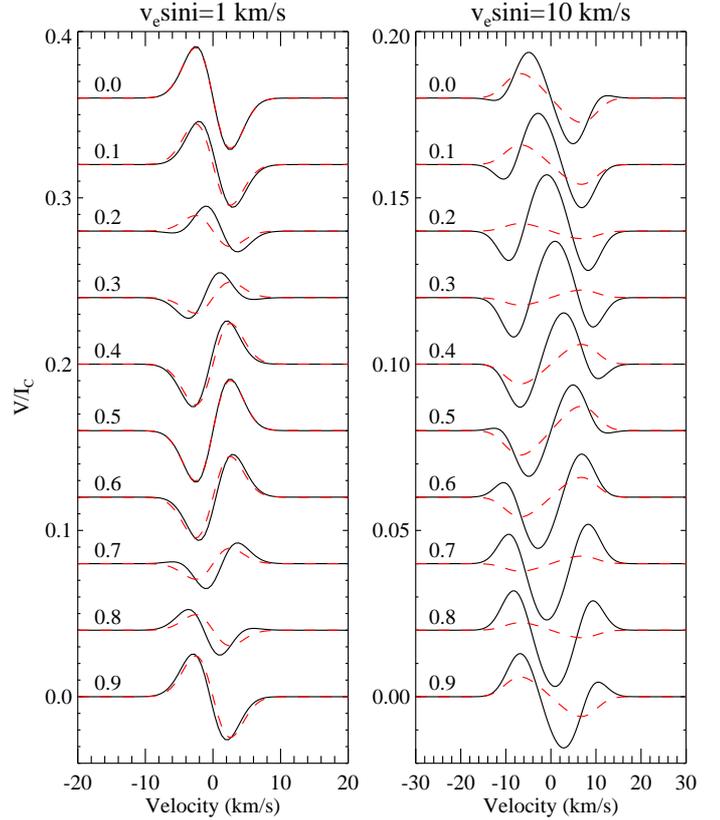}
\caption{Comparison of the Stokes $V$ signatures for the dipolar field topology with $i=\beta=90$\degr\ (\textit{solid lines}) and for a homogeneous radial field distribution (\textit{dashed lines}) scaled to give the same mean longitudinal field at each rotational phase. The left and right panels show results for \vs\,=\,1 and 10~\kms, respectively. The Stokes $V$ spectra are offset vertically according to the stellar rotational phase.}
\label{fig:stokesV_model}
\end{figure}

Although not as realistic as e.g. dipolar field, such schematic field model has the advantage of producing the same basic anti-symmetric Stokes $V$ profile shape independently of the viewing angle. Experimenting with different magnetic field distributions, we found that a uniform radial field provides the most conservative estimate of the circular polarization amplitude for a given \bz. In other words, any other surface magnetic field distribution must produce a stronger Stokes $V$ signature for the same \bz. This is illustrated in Fig.~\ref{fig:stokesV_model}, where we show the Stokes $V$ profiles computed for a perpendicular dipolar field and for radial fields with the same \bz\ at each rotational phase. Evidently, the radial field distributions yield a lower amplitude of the circular polarization signal. This difference is small for very low \vs\ but then quickly increases for larger projected rotational velocities because of the crossover effect \citep{mathys:1995a}. Thus, one can ascertain that the Stokes $V$ profiles corresponding to a homogeneous radial field provide a robust lower limit on the circular polarization signal that must accompany a given \bz\ measurement. 

\begin{table*}
\caption{Atmospheric parameters and rotational velocities of HgMn stars.}
\label{tab1}
\centering
\begin{tabular}{r c c c c c c c c}
\hline\hline
\multicolumn{3}{c}{Star} &  $T_{\rm eff}$ & $\log g$ & Reference & \vs\ & Reference & \vs, this study \\
\multicolumn{1}{c}{HD} & HR & Other & (K) & (cgs) & & (\kms) &  & (\kms) \\
\hline
35548  & 1800 &  & 11100 & 3.80 & 1 & 1.0--2.0 & 1, 2 & $1.0\pm0.2$ \\
65949  &    &  & 13100 & 4.00 & 3 & 0.0--1.0 & 3 & $1.3\pm0.3$ \\
71066  & 3302 & $\kappa^2$\,Vol & 12000 & 4.10 & 4 & 1.5--2.0 & 2, 4 & $1.7\pm0.2$ \\
78316  & 3623 & $\kappa$\,Cnc & 13200 & 3.70 & 10 & 6.5--7.3 & 1, 6, 11 & $6.5\pm0.3$ \\
175640 & 7143 & & 12000 & 3.95 & 5 & 1.5--2.5 & 1, 5, 6, 8 & $1.6\pm0.3$ \\
178065 & 7245 & & 12300 & 3.60 & 6 & 1.5--2.0 & 1, 2, 6, 8  & $1.9\pm0.2$ \\
186122 & 7493 & 46\,Aql & 12560 & 3.80 & 7 & 0.0--2.0 & 1, 2, 6, 7, 8 & $1.1\pm0.2$ \\
193452 & 7775 & $\beta^2$\,Cap & 10750 & 4.00 & 12 & 0.75--2.0 & 1, 2, 6, 8 & $1.2\pm0.2$ \\
209459 & 8404 & 21\,Peg & 10400 & 3.55 & 9 & 3.7--3.8 & 6, 8, 9 & $3.6\pm0.2$ \\
\hline
\end{tabular}
\tablebib{
(1) \citet{dolk:2002}, (2) \citet{hubrig:1999a}, (3) \citet{cowley:2010}, (4) \citet{yuce:2011}, (5) \citet{castelli:2004}, (6) \citet{landstreet:2009}, (7) \citet{castelli:2009}, (8) \citet{hubrig:2001}, (9) \citet{fossati:2009}, (10) \citet{ryabchikova:1998}, (11) \citet{woolf:1999}, (12) \citet{wahlgren:2000}.
}
\end{table*}

\begin{table*}
\caption{Observational data and results of the magnetic field analysis of HgMn stars.}
\label{tab2}
\centering
\begin{tabular}{l c c r c c c}
\hline\hline
\multicolumn{1}{c}{Star} & UT Date & HJD\,$-24\times10^5$ & $T_{\rm exp}$ (s) & $S/N$ & \bz\ (G)\tablefootmark{a} & \bt\ (G)\tablefootmark{b} \\
\hline
  \object{HD\,35548}  & 2010-01-06 & 55202.7401 & 920  & 200 & $-0.7\pm2.5$  & $\le200$ \\ 
  \object{HD\,65949}  & 2011-02-08 & 55600.7080 & 2400 & 180 & $-2.8\pm4.6$  & $\le300$ \\ 
  \object{HD\,71066}  & 2010-01-09 & 55205.7867 & 2000 & 400 & $-1.1\pm0.8$  & $\le200$ \\
  \object{HD\,78316}  & 2010-01-15 & 55211.7873 & 1200 & 370 & $-0.4\pm2.7$ & $\le700$ \\
  \object{HD\,175640} & 2010-05-01 & 55317.8656 & 1200 & 200 & $-0.6\pm2.2$  & $\le250$ \\
  \object{HD\,178065} & 2010-05-03 & 55319.8317 & 1600 & 220 & $-2.3\pm2.1$  & $\le500$ \\
  \object{HD\,186122} & 2010-05-03 & 55319.8505 & 1400 & 230 & $\phantom{-}0.7\pm1.9$  & $\le200$ \\
  \object{HD\,193452} & 2010-05-03 & 55319.8683 & 1200 & 240 & $-1.0\pm1.4$  & $\le250$ \\
  \object{HD\,209459} & 2010-08-09 & 55417.6295 & 1600 & 90 & $-2.2\pm6.0$  &  \\
                                  & 2010-08-13 & 55421.7623 & 800 & 170 & $-4.6\pm3.1$  & $\le600$ \\
\hline
\end{tabular}
\tablefoot{
\tablefoottext{a}{Longitudinal magnetic field from \citet{makaganiuk:2011a} except for HD\,65949 and HD\,209459.}
\tablefoottext{b}{Upper limit on the turbulent magnetic field strength at the 95\% confidence level.}
}
\end{table*}

\section{Results}
\label{results}

\subsection{Non-detection of the longitudinal magnetic field in HD\,65949 and HD\,209459}
\label{nobz}

\begin{figure}[!th]
\centering
\figps{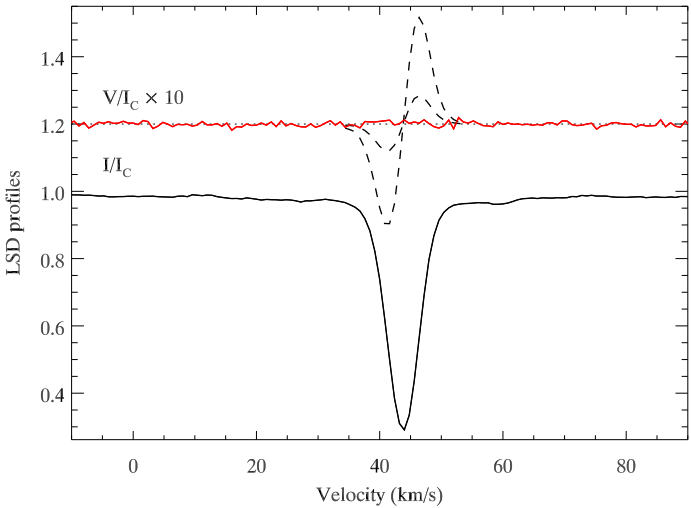}
\caption{LSD Stokes $I$ and $V$ profiles for HD\,65949. The mean polarization profile is offset vertically and expanded by a factor of 10 relative to Stokes $I$. The dashed lines show the circular polarization signatures for \bz\,=\,$-77$ and $-290$~G, corresponding to the range of the FORS1/2 longitudinal field detections reported by \citet{hubrig:2012}.}
\label{fig:lsd65949}
\end{figure}

The outcome of our single Stokes $V$ observation of the extreme HgMn star HD\,65949 \citep{cowley:2010} is illustrated in Fig.~\ref{fig:lsd65949}. The LSD Stokes $V$ profile has no polarization signature stronger than $10^{-3}$. The mean longitudinal magnetic field, \bz\,=\,$-2.8\pm4.6$~G, inferred from the first moment of the circular polarization profile is also formally consistent with zero. Thus, we conclude that there is no evidence of a magnetic field on HD\,65949. 

\citet{hubrig:2012} published five FORS1/2 longitudinal magnetic field measurements of HD\,65949 with typical error bars of 20--60~G. Four of their \bz\ values, distributed over 7 years, suggest a magnetic field detection in the range $-77$~G to $-290$~G. \citet{bagnulo:2012} reanalyzed the FORS1 observation from which the last (and the strongest) of these measurements was derived, finding \bz\,=\,$-42\pm79$~G. They conclude that in this and many other cases, the FORS detections of weak magnetic fields are not reliable.

We used the method described in Sect.~\ref{vmodel} to predict the Stokes $V$ signatures corresponding to the \bz\ values reported by \citet{hubrig:2012}. As demonstrated by the dashed lines in Fig.~\ref{fig:lsd65949}, a longitudinal field at the level of $-77$ to $-290$~G should produce a prominent polarization signature, readily detectable with the LSD profile quality achieved in our observations. It is highly improbable that this huge difference between the predicted and observed polarization signature is due to rotational modulation. Thus, the HARPSpol data are incompatible with previous claims of the longitudinal magnetic field detections in HD\,65949.

\begin{figure}[!th]
\centering
\figps{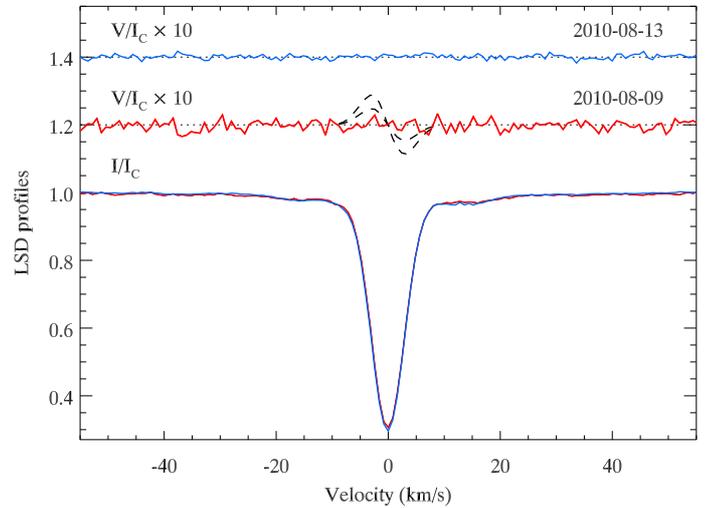}
\caption{LSD Stokes $I$ and $V$ profiles for the two HARPSpol observations of HD\,209459. The mean polarization profiles are offset vertically and expanded by a factor of 10 relative to Stokes $I$. The dashed lines show the circular polarization signatures for \bz\,=\,$53$ and $99$~G, corresponding to the range of the longitudinal field detections claimed by \citet{hubrig:2012} from the same observational data.}
\label{fig:lsd209459}
\end{figure}

Two HARPSpol spectropolarimetric observations are available for the superficially normal late-B star HD\,209459 (21~Peg). The LSD Stokes $I$ and $V$ profiles derived from these spectra are presented in Fig.~\ref{fig:lsd209459}. No evidence of the Zeeman Stokes $V$ signatures can be found above the noise level of $1.7\times10^{-3}$ for the first (9 Aug 2010) and $8.7\times10^{-4}$ for the second (13 Aug 2010) observation. The corresponding longitudinal magnetic field is $-2.2\pm6.0$~G and $-4.6\pm3.1$~G, respectively. Thus, it is unlikely that 21~Peg possesses a large-scale field stronger than a few G. 

Contrary to these results, \citet{hubrig:2012} reported a 50--100~G longitudinal field based on the moment technique analysis  \citep{mathys:1991} of the same HARPSpol spectra of 21~Peg. Magnetic field was presumably found in the lines of Ti, Cr, and Fe, which comprise 80\% of the features included in our LSD line mask for this star. Surprisingly, these detections were reported only for the lower quality spectrum but not for the higher quality one obtained four days later. The minimum Stokes $V$ signature corresponding to the \bz\ measurements reported by \citet{hubrig:2012} significantly exceeds the noise level of the LSD Stokes $V$ profiles (see Fig.~\ref{fig:lsd209459}). The lower S/N observation is incompatible with the predicted polarization signature for \bz\,=\,53~G at the confidence level of $\gg$\,99.9\% according to a $\chi^2$ probability analysis. The lack of polarization signal in the data is therefore grossly inconsistent with the longitudinal field detections obtained for 21~Peg using the moment technique.

\subsection{Magnetic intensification of the Fe~{\sc ii} 6147 and 6149 \AA\ spectral lines}
\label{intens}

The singly ionized iron absorption lines at 6147.7 and 6149.3~\AA\ share the same upper atomic energy level and have nearly identical oscillator strengths. However, their Zeeman splitting patterns are very different: the 6147.7~\AA\ line splits in two $\pi$ and four $\sigma$ components whereas the 6149.3~\AA\ line has two $\sigma$ components coinciding in wavelength with the pair of $\pi$ components. As a result, the latter line exhibits a simple doublet Zeeman splitting pattern, which makes this line particularly useful for direct measurements of the mean magnetic field modulus in slowly rotating Ap stars \citep{mathys:1997b}, despite its relatively small effective Land\'e factor ($z=1.35$). 

The two \ion{Fe}{ii} lines have very similar intensity in normal stars. On the other hand, in the presence of the magnetic field with a strength above a few kG, the \ion{Fe}{ii} 6147.7~\AA\ line experiences a stronger magnetic intensification compared to the 6149.3~\AA\ line. \citet{mathys:1990} and \citet{mathys:1992} noted that in the field strength interval from 3 to 5 kG the normalized equivalent width difference of these lines, $\Delta W_{\lambda}/\langle W_{\lambda}\rangle\equiv 2(W_{6147}-W_{6149})/(W_{6147}+W_{6149})$, follows an approximately linear trend with mean magnetic field modulus. \citet{mathys:1992} warned against extrapolating this relation to weaker fields. Nevertheless, several subsequent studies of Am and HgMn stars ascribed 5--10\% intensification of 6147.7 relative to 6149.3~\AA\ to the presence of 2--3 kG magnetic fields with a complex structure \citep{lanz:1993a,hubrig:1999a,hubrig:2001}.

A handful of Ap stars have fields stronger than $\sim$\,5~kG, causing the \ion{Fe}{ii} 6147.7 and 6149.3~\AA\ lines to exhibit incomplete Paschen-Back splitting \citep{mathys:1990}, which has been investigated in detail \citep{takeda:1991,stift:2008}. The formation of these lines in weak magnetic fields has not been studied as thoroughly. \citet{takeda:1991} carried out theoretical disk-center line profile and equivalent width calculations neglecting magneto-optical effects. Part of his conclusions were based on a simplified analytical treatment of polarized radiative transfer (PRT) in a Milne-Eddington atmosphere. Subsequently, \citet{nielsen:2002} presented another set of differential intensification calculations based on mimicking PRT by adding individual Zeeman components to the input line list for regular unpolarized spectrum synthesis. Both studies suggested significant non-linearities in the weak-field behavior of the magnetic intensification. 

\begin{figure}[!th]
\figps{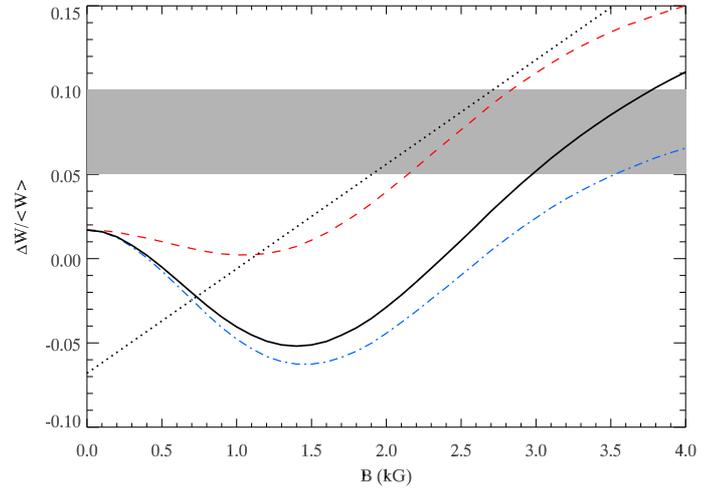}
\caption{Relative intensification of the \ion{Fe}{ii} 6147.7 and 6149.3~\AA\ lines as a function of the magnetic field strength. Different curves show results for turbulent (\textit{solid line}), radial (\textit{dashed line}), and azimuthal (\textit{dash-dotted line}) magnetic field. The dotted line shows an empirical relation for strong-field Ap stars \citep{mathys:1992}. The shaded rectangular region corresponds to the intensification measurements reported for HgMn stars in the literature.}
\label{fig:ew}
\end{figure}

Here we investigated behavior of the \ion{Fe}{ii} 6147.7 and 6149.3~\AA\ lines for different magnetic field geometries using accurate numerical PRT calculations discussed in Sect.~\ref{imodel}. Adopting the oscillator strengths $\log gf=-2.827$ and $-2.841$ \citep{raassen:1998} for the 6147.7 and 6149.3~\AA\ line, respectively, we computed the relative intensification factors and examined the resulting line profiles for the 0--4~kG magnetic field range. These calculations took into account the incomplete Paschen-Back effect in the \ion{Fe}{ii} lines, although deviations from the linear Zeeman splitting are relatively small in the range of field strengths considered here. 

Figure~\ref{fig:ew} shows $\Delta W_{\lambda}/\langle W_{\lambda}\rangle$ as a function of the magnetic field strength for the homogeneous radial, azimuthal, and turbulent fields. Evidently, the weak-field behavior of the \ion{Fe}{ii} line pair is complex and does not follow a linear extrapolation from the strong-field regime. In the field interval 1--2~kG significant ``negative'' intensification is expected for the azimuthal and turbulent magnetic fields. Formally, the same intensification factor is obtained in the absence of the field and for fields as strong as 1.5--3.0~kG. This means that comparing only the equivalent widths of these two \ion{Fe}{ii} lines will not yield an unambiguous detection of a magnetic field below 2.5--3.0~kG because their relative intensification curve is strongly non-linear in this regime and depends on an \textit{a priori} unknown field topology.

\begin{figure}[!th]
\figps{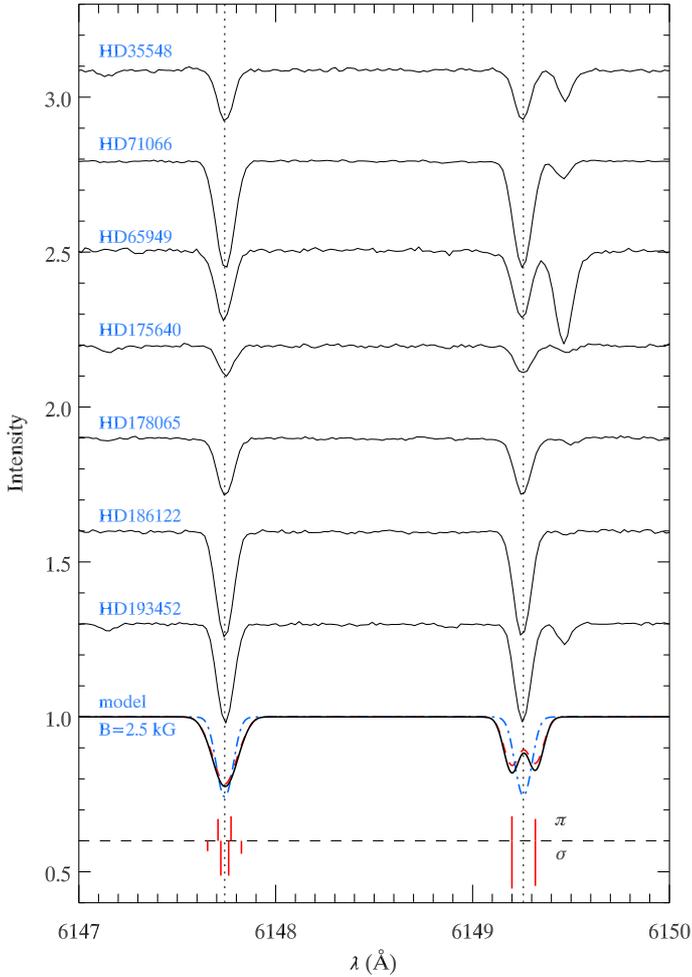}
\caption{Profiles of the \ion{Fe}{ii} 6147.7 and 6149.3~\AA\ lines in the spectra of HgMn stars. The observed spectra are offset vertically for clarity. The unshifted spectra show theoretical calculations for 2.5~kG radial (\textit{dashed line}) and turbulent (\textit{solid line}) magnetic field. The dash-dotted line corresponds to non-magnetic theoretical spectrum. The vertical dotted lines indicate central wavelengths of the \ion{Fe}{ii} lines. The bottom plot schematically illustrates the Zeeman splitting patterns of the two \ion{Fe}{ii} lines.}
\label{fig:prf}
\end{figure}

In previous studies of HgMn stars, the intensification factors $\Delta W_{\lambda}/\langle W_{\lambda}\rangle$ at the level of 5--10\% were attributed to the presence of magnetic field. Figure~\ref{fig:ew} shows that this interpretation requires the mean field intensity to be above 2~kG for radial and above 3~kG for turbulent and azimuthal fields, respectively. Our calculations show that for such fields the Zeeman broadening and splitting of spectral lines is considerable and should be easily detectable with $R\ge10^5$ observations of sharp-lined HgMn stars. This is illustrated in Fig.~\ref{fig:prf}, which compares the observed spectra of the HgMn stars from our sample with the PRT calculations for the stellar parameters \teff\,=\,12000~K, \lgg\,=\,4.0, $[Fe]=+0.5$, $B=2.5$~kG, and for \vs\,=\,1.5~\kms\ typical of our target stars. Independently of the magnetic field geometry, a noticeable broadening of the \ion{Fe}{ii} 6147.7~\AA\ line and a splitting of the 6149.3~\AA\ line occurs for the magnetic field intensity exceeding 1.5--2.0~kG. But the actual high-resolution observed spectra of the HgMn stars analyzed here and presented in previous studies (e.g. see Fig.~1 in \citealt{hubrig:2001}) conspicuously lack any signs of the magnetically-induced spectral line profile distortions. The lack of such distortions rules out field moduli above 1.5--2.0~kG.

Figure~\ref{fig:prf} also shows that the \ion{Hg}{ii} 6149.5~\AA\ line is present in most of our stars and is particularly prominent HD\,65949 due to its high mercury overabundance. For stars with \vs\ exceeding a few \kms\ this feature will blend the \ion{Fe}{ii} 6149.3 line, compromising the equivalent width analysis \citep[see also][]{takada-hidai:1992}.

\begin{figure}[!th]
\centering
\figps{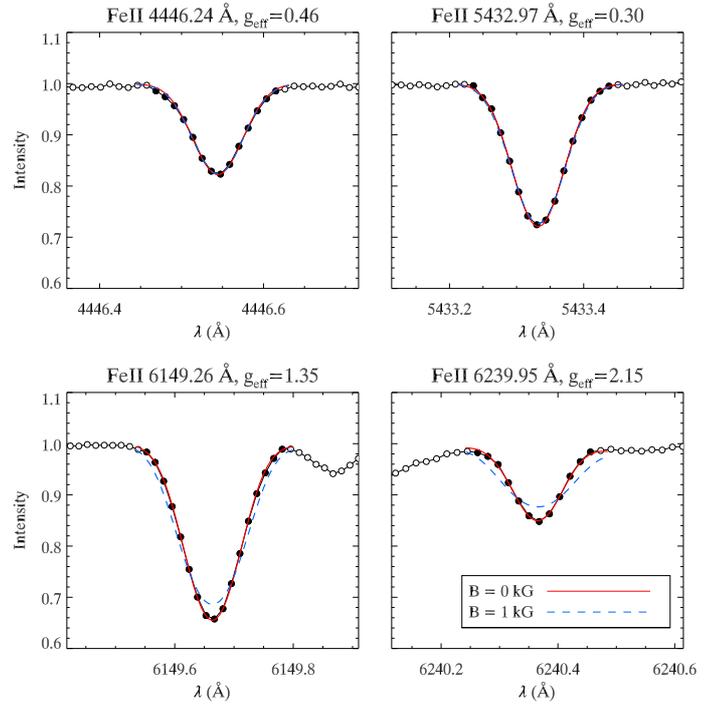}
\caption{Comparison of observations and synthetic spectra computed for \ion{Fe}{ii} lines of different magnetic sensitivity in the spectrum of HD\,71066. Observations are shown with symbols. Filled circles indicate the wavelength interval considered for the least-squares fit. The solid line represents synthetic spectrum in the absence of magnetic field. The dashed line corresponds to the calculations for 1 kG turbulent field.}
\label{fig:lfit}
\end{figure}

Thus, we conclude that magnetic field detections based on comparing the equivalent widths of the \ion{Fe}{ii} 6147.7 and 6149.3~\AA\ lines are ambiguous and potentially misleading. In fact, for the slowly rotating stars, magnetic intensification is less informative than the profile shapes of the same spectral lines. The observed \ion{Fe}{ii} line profiles of all 7 HgMn stars with \vs\,$\le$\,2~\kms\ are incompatible with previous studies that interpreted 6147.7 vs. 6149.3~\AA\ equivalent widths in terms of magnetic fields.

\subsection{Upper limits on tangled magnetic fields}
\label{bturb}

The Zeeman broadening analysis of each target star was carried out in several steps. First, we calculated a grid of synthetic spectra for different turbulent magnetic field strengths covering the entire HARPS wavelength range. For these calculations we adopted the model atmosphere parameters listed in Table~\ref{tab1} and compiled information on stellar abundances from previous studies. The atomic line data were extracted from the VALD database. Using this grid of synthetic spectra, we identified two groups of unblended spectral lines with a different response to magnetic field. We used the first group of about 10--15 spectral lines with small effective Land\'e factors (typically $z\le0.5$) to determine \vs. The central wavelengths and Land\'e factors of these lines are given in Table~\ref{tab3}. We used a least-squares routine to fit each spectral line individually, allowing stellar radial velocity and elemental abundance to vary. The last column of Table~\ref{tab1} reports the mean and standard deviation of our projected rotational velocities.

For the two SB2 stars in our sample, HD\,35548 and HD\,78316 ($\kappa$\,Cnc), special effort was made to avoid spectral regions affected by the absorption lines of the companion. We also corrected synthetic spectra for the continuum dilution by the secondary stars using the light ratios derived by \citet{ryabchikova:1998} and \citet{dolk:2002} for HD\,78316 and HD\,35548, respectively.

\begin{table}
\caption{Spectral lines used for broadening analysis.}
\label{tab3}
\centering
\begin{tabular}{c c c | c c c}
\hline\hline
Ion & $\lambda$ (\AA) & $z$ &  Ion & $\lambda$ (\AA) & $z$ \\
\hline
\multicolumn{6}{c}{low-$z$ lines}\\
\ion{Fe}{ii}  & 4278.159 &  0.148 &  \ion{Fe}{ii} &  4893.820 &  0.386 \\ 
\ion{Cr}{ii}  & 4284.188 &  0.324 &  \ion{P}{ii}  &  4954.367 &  0.500 \\
\ion{Fe}{ii}  & 4296.572 &  0.583 &  \ion{Fe}{ii} &  4993.358 &  0.627 \\
\ion{Fe}{ii}  & 4314.310 &  0.354 &  \ion{Fe}{ii} &  5006.841 &  0.609 \\
\ion{Fe}{ii}  & 4369.411 &$-0.120$&  \ion{Fe}{ii} &  5007.447 &  0.500 \\
\ion{Fe}{ii}  & 4384.319 &  0.670 &  \ion{Fe}{ii} &  5029.097 &  0.215 \\
\ion{Fe}{ii}  & 4446.237 &  0.458 &  \ion{Fe}{ii} &  5082.230 &  0.548 \\
\ion{Ti}{ii}  & 4464.448 &  0.492 &  \ion{Fe}{ii} &  5098.685 &  0.181 \\
\ion{Fe}{ii}  & 4491.405 &  0.420 &  \ion{Fe}{ii} &  5143.880 &  0.484 \\
\ion{Fe}{ii}  & 4508.288 &  0.505 &  \ion{Fe}{ii} &  5149.465 &  0.539 \\
\ion{Cr}{ii}  & 4634.070 &  0.530 &  \ion{Fe}{ii} &  5197.577 &  0.667 \\
\ion{Ti}{ii}  & 4657.200 &  0.588 &  \ion{Fe}{ii} &  5237.950 &  0.512 \\
\ion{Fe}{ii}  & 4663.708 &  0.404 &  \ion{Fe}{ii} &  5432.967 &  0.299 \\
\ion{Fe}{ii}  & 4731.453 &  0.651 &  \ion{Fe}{ii} &  5498.576 &  0.461 \\
\ion{Ti}{ii}  & 4763.881 &  0.348 &  \ion{Fe}{ii} &  5534.847 &  0.573 \\
\ion{Ti}{ii}  & 4798.521 &  0.398 &  \ion{P}{ii}  &  6034.039 &  0.500 \\
\multicolumn{6}{c}{high-$z$ lines}\\
\ion{Fe}{ii} & 4263.869 &  1.940 &  \ion{Fe}{ii} &  5450.099 &  2.806 \\
\ion{Fe}{ii} & 4273.326 &  2.155 &  \ion{Fe}{ii} &  5737.898 &  2.244 \\
\ion{Ti}{ii} & 4320.950 &  2.175 &  \ion{Fe}{ii} &  5830.341 &  2.282 \\
\ion{Fe}{ii} & 4385.387 &  1.330 &  \ion{Fe}{ii} &  6149.258 &  1.350 \\
\ion{Fe}{ii} & 4461.439 &  1.725 &  \ion{Fe}{ii} &  6239.953 &  2.150 \\
\ion{Fe}{ii} & 4580.063 &  1.850 &  \ion{Fe}{ii} &  6432.680 &  1.825 \\
\hline
\end{tabular}
\end{table}

Next we analyzed in detail a smaller (5--10) set of magnetically sensitive spectral lines (see Table~\ref{tab3}). Most of them had $z\ge1.5$, but a few, like \ion{Fe}{ii} 4385.387~\AA\ and \ion{Fe}{ii} 6149.3~\AA, provided useful constraints on the magnetic field despite a relatively small Land\'e factor, due to their unusual Zeeman splitting patterns. The list of usable spectral lines was dominated by \ion{Fe}{ii}, but varied from star to star depending on the chemical composition, blending, and the quality of observations. Several magnetically sensitive lines, notably \ion{Fe}{ii} 4273.3 and 6149.3~\AA, were analyzed in each star. We avoided very strong \ion{Fe}{ii} lines, such as 4923.9 and 5018.4~\AA, which might be affected by NLTE effects in the line core.

For each magnetically sensitive line we performed a series of least-squares fits for different \vs\ and turbulent magnetic field strengths. These parameters were varied in steps of 0.1~\kms\ and 50~G within $\pm1$~\kms\ around the previously determined \vs\ and for \bt\ between 0 and 1000~G, respectively. Since we considered only the magnetic distortion of the spectral line shapes, the element abundance was adjusted individually for each \vs-\bt\ pair. 

Fig.~\ref{fig:lfit} presents examples of our spectrum synthesis fits for different magnetic field strengths. It shows a pair of low-$z$ \ion{Fe}{ii} lines together with two magnetically sensitive lines in the spectrum of HD\,71066. One can see that, after adjusting the line strength, it is possible to obtain an excellent fit for both groups of lines with zero magnetic field intensity. On the other hand, if the field strength is increased to 1~kG, the two low-$z$ lines are not affected while the calculated profiles of the magnetically sensitive lines become broader than the observations.

\begin{figure}[!th]
\centering
\fifps{8cm}{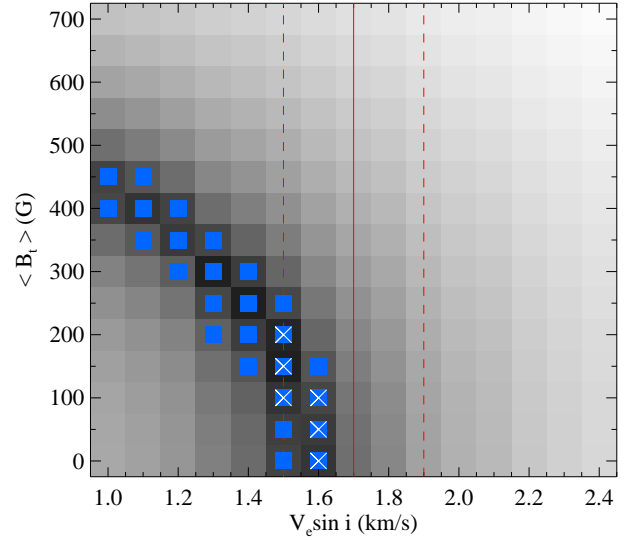}
\caption{Derivation of the upper limit on the turbulent magnetic field strength in HD\,71066 using the \ion{Fe}{ii} 4273.3~\AA\ line. The $\chi^2$ of the fit (background image) is shown as a function of \bt\ and \vs. The squares indicate parameter combinations compatible with observations at the 95\% confidence level. The crosses show \bt-\vs\ pairs consistent with the projected rotational velocity (vertical lines) determined from the magnetically insensitive spectral lines.}
\label{fig:chi4273}
\end{figure}

\begin{figure}[!th]
\centering
\fifps{8cm}{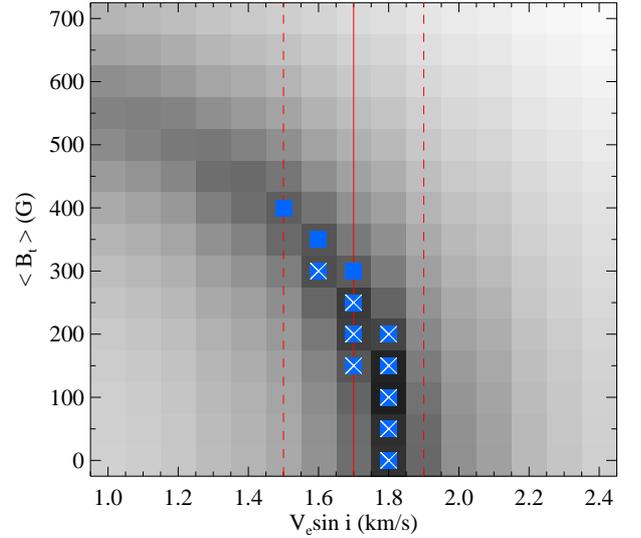}
\caption{Same as Fig.~\ref{fig:chi4273}, but for the \ion{Fe}{ii} 6149.3~\AA\ line.}
\label{fig:chi6149}
\end{figure}

To obtain quantitative constraints on the turbulent magnetic field strength we examined the fit $\chi^2$ as a function of \vs\ and \bt, as illustrated in Fig.~\ref{fig:chi4273}. The confidence level of the fit was characterized by the $\chi^2$ probability function, $P_{\chi^2}$. For a typical spectral line with a triplet-like Zeeman splitting, the effects of a weak magnetic field and rotational Doppler broadening largely compensate each other, so an adequate fit can be achieved with different combinations of \vs\ and \bt. This degeneracy can be lifted by considering the \vs\ derived from magnetically insensitive lines. Assuming that the probability of finding a particular value of \vs\ follows a Gaussian distribution, $P_{\rm g}$, with the center and width given by the mean \vs\ and the corresponding error (see Table~\ref{tab1}), we constructed the total probability function $P_{\rm tot} (v_{\rm e}\sin i, \langle B_{\rm t} \rangle) = P_{\chi^2} P_{\rm g}$. In this way, we combined information about the fit quality for a particular magnetically sensitive line with the \vs\ constraint from lines not affected by magnetic field. 

A few spectral lines have Zeeman splitting patterns that are wide doublets, making it possible to distinguish magnetic broadening from rotational Doppler broadening. This allowed us to disentangle \vs\ and \bt\ even for a relatively low magnetic field strength. Fig.~\ref{fig:chi6149} provides an example of this situation for the \ion{Fe}{ii} 6149.3~\AA\ line in the spectrum of HD\,71066. In this case the observations are clearly not compatible with the slow rotation/strong field scenario because the Zeeman splitting cannot mimic rotational broadening.

We analyzed 8 HgMn and one normal late-B star. None of these stars require the presence of a magnetic field to match the observed profiles of spectral lines with different magnetic sensitivity. Adopting a confidence level of 95\%, we inferred upper limits for the turbulent magnetic field in each star, using 2--3 spectral lines that provide the best constraint. These upper limits are reported in the last column of Table~\ref{tab2}. Typically, we rule out fields above 200--300~G for the HgMn stars with low projected rotational velocities (\vs\,$\le$\,2~\kms). A higher limit of 500~G was obtained for one sharp-lined HgMn star, HD\,178065 because low iron abundance weakens many useful \ion{Fe}{ii} diagnostic features (see Fig.~\ref{fig:prf}). For the two more rapidly rotating stars, HD\,78316 ($\kappa$\,Cnc) and HD\,209459 (21\,Peg), we constrained \bt\ to be below 600--700~G. Therefore, the principal conclusion of this section is that \textit{the line broadening in high-resolution spectra of the target stars is incompatible with the idea of ubiquitous presence of multi-kG complex magnetic fields} in the HgMn-star atmospheres.

\section{Discussion}
\label{discussion}

\subsection{Tangled magnetic fields in HgMn stars}

Our detailed spectrum synthesis analysis shows that modern $R>10^5$ spectra could detect the weak magnetic broadening that would be caused by 1--2~kG unstructured fields in slowly rotating HgMn stars. However, observed profiles of lines with different magnetic sensitivity rule out tangled fields stronger than a few hundred G in the atmospheres of the HgMn stars studied here.

We showed that some claims of strong complex fields in HgMn stars, in particular those that interpret equivalent widths of \ion{Fe}{ii} 6147.7 and 6149.3~\AA\ in terms of differential magnetic intensification, relied on incorrect assumptions about spectral line formation and therefore cannot be substantiated.

Our magnetic line formation calculations suggest that unambiguous detection of magnetic intensification with this \ion{Fe}{ii} line pair requires a field so strong that Zeeman splitting would be obvious in slowly rotating stars. Of course, the latter effect can be masked by line broadening in faster rotators, but in such stars the differential intensification method is not reliable anyway due to line blending problems \citep{takada-hidai:1992}.

Highly sensitive analyses of the Stokes $V$ LSD profiles of HgMn stars \citep[see Sect.~\ref{nobz} and ][]{makaganiuk:2011a} place stringent upper limits on the mean longitudinal magnetic field. The non-detection of the circular polarization signals in high-resolution spectra also rules out the presence of weak complex magnetic fields topologically similar to those found in active late-type stars because the signatures of latter fields are readily detectable with HARPSpol and other modern spectropolarimeters \citep[e.g.][]{kochukhov:2011}. Thus, it should be realized that the putative tangled fields of HgMn stars must be substantially weaker and/or more complex than the dynamo fields of cool stars to evade detections with modern high-resolution spectropolarimeters.

Recently, weak sub-G fields have been detected on Vega and Sirius \citep{petit:2010,petit:2011}. We cannot rule out such fields on HgMn stars or any other type of intermediate-mass stars. However, such fields would not be dynamically important, and therefore are not capable of providing an explanation for the chemical spot formation in HgMn stars. Thus, for the purpose of understanding the dichotomy between non-magnetic (HgMn, Am) and strongly magnetic (Ap/Bp) intermediate-mass stars, we find no compelling reasons to revise the ``non-magnetic'' classification of HgMn stars.

\subsection{Longitudinal field detections for HgMn stars}

Currently, the most precise constraints on magnetic fields in HgMn stars are obtained by applying the LSD technique to high-resolution circular polarization spectra \citep{auriere:2010a,makaganiuk:2011a,kochukhov:2011b}. However, the null results of the HARPSpol magnetic survey of HgMn stars \citep{makaganiuk:2011a} and non-detections of the magnetic field in individual objects \citep{makaganiuk:2011,makaganiuk:2012} analyzed with the LSD method were recently challenged by \citet{hubrig:2012}. These authors re-analyzed the same high-resolution HARPSpol polarization spectra with an alternative moment technique \citep{mathys:1991} and in a few cases reported 3--4$\sigma$ detections of 30--100~G longitudinal magnetic fields. Analysis was carried out for individual chemical elements, resulting in \bz\ error bars 10--30~G, significantly exceeding those obtained for the same targets by the LSD approach.

\citet{hubrig:2010,hubrig:2012} invoked inhomogeneous distributions of chemical elements to explain why the moment analysis of individual elements yields much stronger fields than permitted by the LSD results. These authors suggested that the (unspecified) distribution of spots on HgMn stars enhances the polarization signal for inhomogeneously distributed elements but suppresses the signal for Fe-peak elements that dominate the LSD line mask. However, in the study of the HgMn star HD\,11753 \citet{makaganiuk:2012} applied LSD to individual elements, including those with the strongest horizontal abundance gradients, without detecting a field at any epoch. Furthermore, results obtained by \citet{hubrig:2012} for this star seem to contradict their hypothesis because the longitudinal field is fairly coherent for all analyzed elements, including Fe, at the two phases with claimed magnetic field detections. In addition, discrepancies between the application of the moment method and the LSD polarization analyses were also found for non-peculiar stars without chemical spots, e.g. for 21~Peg (Sect.~\ref{nobz}) and for pulsating B-type stars \citep{shultz:2012}.

\begin{figure}[!th]
\centering
\fifps{8cm}{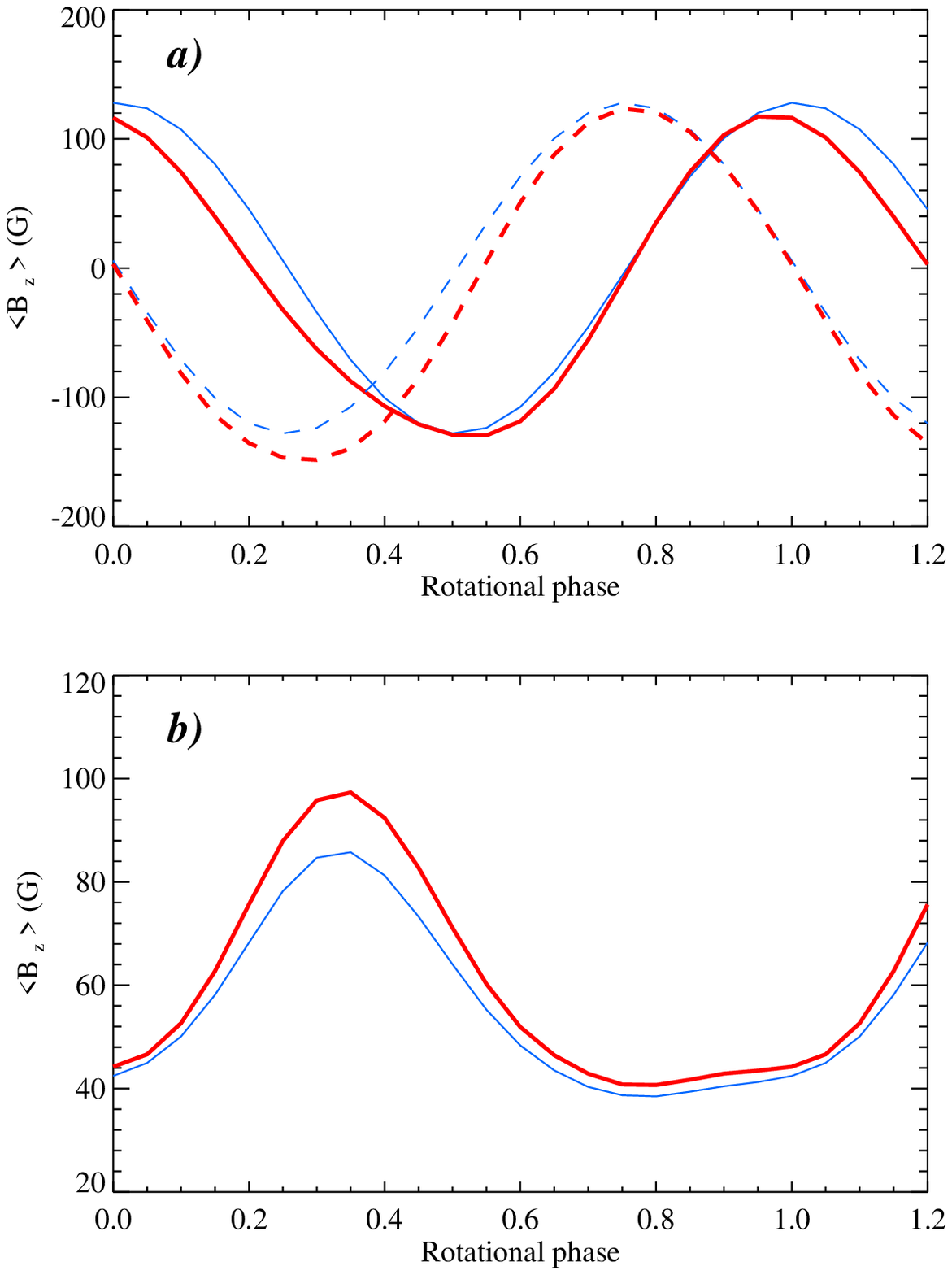}
\caption{Effect of Y spots on the determination of the mean longitudinal magnetic field of the HgMn star HD\,11753. The curves show \bz\ as a function of rotation phase for the Y map derived by \citet{makaganiuk:2012} (\textit{thick lines}) and for a uniform element distribution (\textit{thin lines}). {\bf a)} calculations for the dipolar field topology with a polar strength of 500~G, obliquity $\beta=90\degr$, and two different orientations with respect to the reference rotation phase (\textit{solid and dashed lines}). {\bf b)} calculations for the radial magnetic field map scaled according to the Y abundance distribution.}
\label{fig:Y_model}
\end{figure}

In general, one does not expect topologically simple and low-contrast chemical spots on HgMn stars to have a large effect on the longitudinal magnetic field measurements. Using results from a Doppler imaging analysis of HD\,11753 \citep{makaganiuk:2012}, we assessed in detail the influence of inhomogeneous Y distribution on \bz\ measurements obtained from the spectral lines of this element. Theoretical Stokes $I$ and $V$ spectra were computed using the {\sc Invers10} code \citep{piskunov:2002a} for the Y map derived by \citet{makaganiuk:2012} and for a homogeneous distribution of this element. For these calculations we adopted the dipolar field topology with a polar strength $B_{\rm p}=500$~G, obliquity $\beta=90\degr$, and two different azimuth angles, $\varphi=0$\degr\ and 90\degr. The resulting \bz\ curves obtained from the synthetic spectra are presented in Fig.~\ref{fig:Y_model}a. It is evident that even for Y, which shows the largest abundance contrast across the surface of HD\,11753, the non-uniform element distribution results in no more than 10--20\% change of \bz. Depending on the field geometry, one may expect reduction or amplification of the longitudinal field inferred from the lines of a non-uniformly distributed element. 

In another test calculation we considered an outward-directed radial field with a local intensity that was scaled between 0 and +200~G according to the (logarithmic) abundance of Y. We then compared \bz\ curves inferred from synthetic Y lines computed for this field, assuming either a nonuniform or a homogeneous element distribution. Figure~\ref{fig:Y_model}b demonstrates that with this magnetic field distribution, which perfectly matches the yttrium abundance spots as in the scenarios implied by \citet{hubrig:2010,hubrig:2012}, we find that \bz\ is enhanced by a mere 12\% at the phase when Y line strength is at maximum. These calculations confirm that weak chemical inhomogeneities found on HgMn stars do not significantly enhance magnetic field detections and cannot be responsible for the discrepancies between the LSD and moment technique results.

The LSD technique determines the average Stokes $V$ profile, and not just a single scalar quantity, e.g. the mean longitudinal magnetic field. This allows one to reveal topologically complex or crossover magnetic fields, which yield a detectable Zeeman signature even when the first-order moment of Stokes $V$ is negligible. Simultaneously, one can use various objective statistical methods to assess compatibility of the theoretical magnetic field models with the observed LSD profiles \citep[e.g.][]{shultz:2012,petit:2012}. Here we used this opportunity to directly test the \bz\ moment measurements reported by \citet{hubrig:2012} for HD\,11753 and other HgMn stars.

\begin{figure}[!th]
\centering
\figps{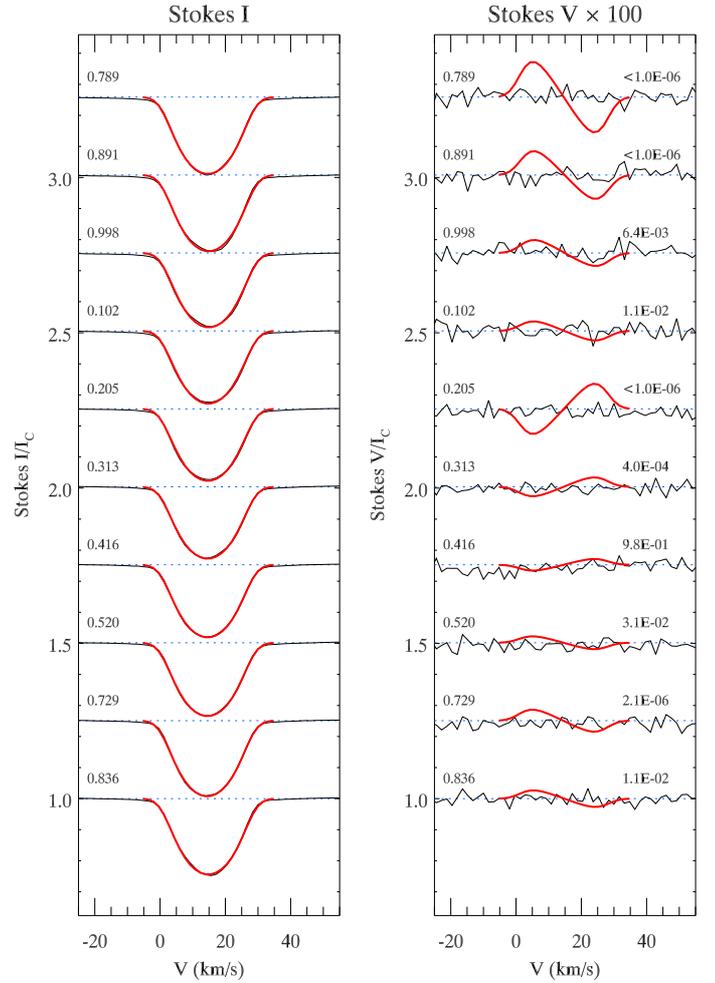}
\caption{LSD Stokes $I$ (\textit{left panel}) and $V$ (\textit{right panel}) profiles of the HgMn star HD\,11753. Thin lines show observations by \citet{makaganiuk:2012} obtained during 10 nights, starting on 2010-01-04 (\textit{top profile}) and ending on 2010-01-14 (\textit{bottom profile}). The thick curves correspond to the minimum-amplitude Stokes $V$ model profiles reproducing the longitudinal field measurements obtained from the same observational data by \citet{hubrig:2012}. Rotational phases are indicated on the left side of each panel. The numbers on the right side of the Stokes $V$ panel report the probability that observations and model profiles are compatible within the error bars.}
\label{fig:lsd_model}
\end{figure}

Using the LSD line profile modeling method described in Sect.~\ref{vmodel}, we have predicted the minimum amplitude of the mean Stokes $V$ profile for the  longitudinal magnetic field measurements published by \citet{hubrig:2012}. The comparison between the observed and modeled Stokes $I$ and $V$ profiles of HD\,11753 is presented in Fig.~\ref{fig:lsd_model} for all 10 rotational phases analyzed in the original paper by \citet{makaganiuk:2012}. The computed Stokes $V$ signatures correspond to the maximum longitudinal field strength reported by \citet{hubrig:2012} for the same set of rotational phases. The Stokes $V$ panel of this figure reports the False Alarm Probability (FAP), characterizing the probability that the difference between the observed and model profiles can be attributed to observational noise. For all but two or three cases, \bz\ inferred by the moment method is incompatible with the observed LSD profiles. In particular, for the two phases for which \citet{hubrig:2012} claimed 3--4$\sigma$ field detections (0.789, 0.205), we find $FAP<10^{-6}$. There are three more phases (0.891, 0.313, 0.729) when simulated signals are highly improbable ($FAP<10^{-3}$). This shows that the high \bz\ values reported for HD\,11753 by \citet{hubrig:2012} are ruled out by the observed LSD profiles. Similar conclusions can be reached for other HgMn stars observed with HARPSpol (41~Eri, 66~Eri, HD\,78316). Applying the same analysis to the LSD Stokes $I$ and $V$ profiles of the binary HgMn star AR\,Aur \citep{folsom:2010}, we established an upper limit, at the 99\% confidence level, of 45~G and 100~G on the absolute value of \bz\ for the primary and secondary components, respectively. This agrees well with the non-detection of the longitudinal field by \citet{folsom:2010} but is grossly incompatible with $|\langle B_{\rm z} \rangle|$ values of up to 560~G for the primary and up to 260~G for the secondary claimed by \citet{hubrig:2010,hubrig:2012}.

We emphasize that our models of the LSD profiles of HD\,11753 and other HgMn stars assume a homogeneous radial field distribution, which corresponds to the smallest possible amplitude of the circular polarization signature for a given \bz. In the presence of stellar rotation with \vs\ above a few \kms, any realistic global field configuration, e.g. a dipole, and especially any complex magnetic field distribution will inevitably lead to additional lobes in the Stokes $V$ profile, yielding a much higher peak-to-peak amplitude of the Zeeman signatures for the same \bz. This would make the discrepancy between predicted Stokes $V$ signatures and the actual observations even more dramatic.

What could be the reason for these discrepant longitudinal field measurements, which are often obtained from the same observational material? Part of the disagreement may be related to different reduction methods. \citet{hubrig:2012} analyzed spectra from a generic HARPS pipeline, whereas we analyzed spectra from the {\sc reduce} code \citep{piskunov:2002} with optimizations designed to yield precise spectropolarimetry. The essential difference between the two reduction procedures is that the polarimetric version of the HARPS pipeline does not include continuum normalization nor the response function correction necessary to compensate for large differences in HARPS fiber throughput. Omitting these steps may lead to a non-negligible spurious continuum polarization. Combined with the longitudinal field measurement technique based on performing a linear least-squares fit (forced through the origin) of the first moment of Stokes $V$ profiles \citep{mathys:1991}, this may result in spurious magnetic field detections.

There could also exist other methodological problems with the applications of the moment technique to weak polarimetric signals. The LSD method has a well-established track record for detecting and characterizing strong \citep{wade:2000}, intermediate \citep{auriere:2007} and extremely weak \citep{petit:2010} stellar magnetic fields. It was also thoroughly tested using theoretical polarized spectrum synthesis calculations for a wide range of magnetic field strengths and orientations \citep{kochukhov:2010a}. In contrast, the moment technique has been systematically applied only to moderate-quality observations of the strong-field Ap/Bp stars \citep{mathys:1991,mathys:1997a}. It is not immediately obvious that an extension of the moment measurements to very low-amplitude circular polarization signals hidden in the noise is robust, especially when it comes to 3--4$\sigma$ \bz\ measurements that are not accompanied by a definite detection of Stokes $V$ signatures. 

To this end, the experience with the FORS low-resolution field detection technique is particularly instructive. Measurements using standard FORS pipeline yielded reasonable results for strong-field Ap stars \citep{kochukhov:2006,landstreet:2007} but, at the same time, suggested detection of weak fields in a variety of objects not known to be magnetic \citep{hubrig:2009}, including HgMn stars \citep{hubrig:2006b}. Very few of these detections were validated by other instruments \citep{silvester:2009,shultz:2012}. These problematic FORS results were eventually traced to poorly documented subtle details of the data reduction and processing, which led to a systematic underestimation of the uncertainty of derived magnetic field strengths \citep{bagnulo:2012}. The moment technique in the implementation used by \citet{hubrig:2012} appears to suffer a similar pitfall when applied to stars with very weak or absent magnetic fields. Unfortunately, these authors did not provide enough detail to reproduce their analysis step by step.

Based on this assessment, we conclude that there is no convincing evidence of the magnetic fields in HgMn stars from high-resolution circular polarization data. Given controversial results and refuted detections, any claim of a significant longitudinal magnetic field in these stars should be considered with caution, if it is not accompanied by the primary magnetic observable: a non-zero Stokes $V$ signature repeatedly detected in individual spectral lines or in mean line profiles.

\subsection{Quadratic fields}

The mean quadratic magnetic field, $\langle B_{\rm q} \rangle \equiv (\langle B^2_{\rm z} \rangle + \langle B^2 \rangle)^{1/2}$, represents an alternative method of extracting information about stellar magnetic fields from the differential broadening of spectral lines with different magnetic sensitivity. This magnetic field diagnostic technique was introduced by \citet{mathys:1995} and further improved by \citet{mathys:2006}. Quadratic field is derived under simplifying assumptions about line formation expressed as multiple linear regressions of observed second-order moments of line profiles against central wavelength, equivalent width, and Zeeman splitting parameters of spectral lines. The quantity $\langle B^2_{\rm z} \rangle + \langle B^2 \rangle$ derived from observations does not separate contributions of the longitudinal field and the field modulus. Due to its attempt to disentangle multiple broadening agents affecting Stokes $I$ spectra, the quadratic field diagnostic method is considerably more complicated to apply compared to, e.g., the mean longitudinal magnetic field measurements. In practice, it is necessary to use different regression relations for different stars and even for different elements in the same star \citep{mathys:2006}.

Quadratic magnetic fields were systematically measured and interpreted with some success for strong-field Ap stars \citep{mathys:1995,landstreet:2000,bagnulo:2002}. At the same time, application of this method to some Ap stars led to occasional controversies since \bq\ was found to be much larger than implied by the separate \bz\ and \bb\ measurements \citep{kupka:1996,kochukhov:2004e}. Furthermore, \citet{kochukhov:2004b} showed that quadratic field can be overestimated by up to 40\% if it is derived from the spectral lines of a non-uniformly distributed chemical element.

Based on application of the quadratic field measurement technique, \citet{mathys:1995b} claimed the discovery of a remarkably strong magnetic field in the HgMn primary of 74\,Aqr. For this star the measurements on two nights gave \bq\,=\,3.6~kG. \citet{hubrig:1998} also reported 2.7~kG quadratic field in $\kappa$~Cnc. More recently, the claims of significant quadratic field detections in HgMn stars were repeated by \citet{hubrig:2012}, who published \bq\ estimates ranging from 1.3~kG in 21\,Peg to 6--8~kG in both components of AR\,Aur. However, no detailed description or illustration of the definitive quadratic field measurements for HgMn stars (as presented by e.g. \citet{mathys:2006} for magnetic Ap stars) can be found in any of these papers. The authors also did not address the biases due to an inhomogeneous element distribution.

In the context of our study, the constraints on turbulent magnetic field strength derived above can be converted to upper limits on the mean quadratic field. For the random-field model considered in our paper, $\langle B^2_{\rm z}\rangle = \langle B^2 \rangle/3$ and $\langle B^2\rangle = \langle B \rangle^2$, so one can find that $\langle B_{\rm q}\rangle \approx 1.15 \langle B_{\rm t}\rangle$. Taking this into account, our results rule out 1.3 and 2.7~kG quadratic fields obtained in previous studies of 21\,Peg and $\kappa$\,Cnc, respectively. No direct comparison can be made for other HgMn stars with the quadratic field detections because their \vs\ are too large for deriving \bt\ with our technique. Nevertheless, we note that detailed spectrum synthesis modeling of the disentangled spectra of the binary components of AR\,Aur carried out by \citet{folsom:2010} does not support the presence of 6--8~kG tangled magnetic fields, since such fields would have resulted in line strengths that were irreproducible by their non-magnetic spectrum synthesis. Similarly, the detailed abundance analysis and non-magnetic spectrum synthesis fits published for 74~Aqr~A by \citet{catanzaro:2006} contradict the presence of 3.6~kG field in this star.

Thus, at this moment, the quadratic magnetic field detections reported for HgMn stars cannot be independently verified and are likely to be spurious given their disagreement with direct line profile analysis. More information about practical details of the \bq\ determination using linear regressions and a comprehensive examination of the limitations of this technique using theoretical spectra is required to assess its reliability and usefulness for HgMn stars.

\begin{acknowledgements}
OK is a Royal Swedish Academy of Sciences Research Fellow, supported by the grants from Knut and Alice Wallenberg Foundation and Swedish Research Council.
\end{acknowledgements}


\end{document}